%% file: conference_101719.tex
\documentclass[conference]{IEEEtran}

\usepackage[dvipsnames]{xcolor}
\IEEEoverridecommandlockouts
\usepackage{cite}
\usepackage{amsmath,amssymb,amsfonts}
\usepackage{algorithmic}
\usepackage{graphicx}
\usepackage{textcomp}
\usepackage{url}
\usepackage{listings}
\usepackage{tikz}
\usepackage{balance}
\usepackage[colorinlistoftodos]{todonotes}

\usepackage{tcolorbox}
\usepackage{subcaption}

\usepackage{pgfplots}
\pgfplotsset{width=\columnwidth,compat=1.9}
\usepgfplotslibrary{external}
\usetikzlibrary{patterns}

\usepackage{comment}
\usepackage[draft]{minted}

\usetikzlibrary{shapes,snakes}
\newcommand*\circled[1]{\tikz[baseline=(char.base)]{
		\node[shape=circle,draw,inner sep=0.5pt] (char) {#1};}}
\usepackage[linesnumbered,vlined,boxed,commentsnumbered, ruled]{algorithm2e} 
\def\BibTeX{{\rm B\kern-.05em{\sc i\kern-.025em b}\kern-.08em
    T\kern-.1667em\lower.7ex\hbox{E}\kern-.125emX}}

\newcommand{\sorrir}{\textsc{Sorrir}}
\newcommand{\bftsmart}{\textsc{BFT-SMaRt}}

\usepackage[absolute,showboxes]{textpos}
\TPMargin{5pt}
\textblockcolour{yellow!00}

\newcommand{\copyrightstatement}{
	\begin{textblock}{0.84}(0.08,0.94)    
		\noindent
		\scriptsize
This is the authors' preprint version of the work. \\
The definitive version is published in the proceedings of the 18th European Dependable Computing Conference, 
12-15 September 2022
Zaragoza, Spain.
	\end{textblock}
}

\makeatletter
\def\PYG@reset{\let\PYG@it=\relax \let\PYG@bf=\relax%
	\let\PYG@ul=\relax \let\PYG@tc=\relax%
	\let\PYG@bc=\relax \let\PYG@ff=\relax}
\def\PYG@tok#1{\csname PYG@tok@#1\endcsname}
\def\PYG@toks#1+{\ifx\relax#1\empty\else%
	\PYG@tok{#1}\expandafter\PYG@toks\fi}
\def\PYG@do#1{\PYG@bc{\PYG@tc{\PYG@ul{%
				\PYG@it{\PYG@bf{\PYG@ff{#1}}}}}}}
\def\PYG#1#2{\PYG@reset\PYG@toks#1+\relax+\PYG@do{#2}}

\@namedef{PYG@tok@w}{\def\PYG@tc##1{\textcolor[rgb]{0.73,0.73,0.73}{##1}}}
\@namedef{PYG@tok@c}{\let\PYG@it=\textit\def\PYG@tc##1{\textcolor[rgb]{0.40,0.40,0.40}{##1}}}
\@namedef{PYG@tok@cp}{\def\PYG@tc##1{\textcolor[rgb]{0.33,0.47,0.60}{##1}}}
\@namedef{PYG@tok@cs}{\let\PYG@bf=\textbf\let\PYG@it=\textit\def\PYG@tc##1{\textcolor[rgb]{0.80,0.00,0.00}{##1}}}
\@namedef{PYG@tok@k}{\let\PYG@bf=\textbf\def\PYG@tc##1{\textcolor[rgb]{0.13,0.53,0.60}{##1}}}
\@namedef{PYG@tok@kp}{\let\PYG@bf=\textbf\def\PYG@tc##1{\textcolor[rgb]{0.00,0.53,1.00}{##1}}}
\@namedef{PYG@tok@kt}{\let\PYG@bf=\textbf\def\PYG@tc##1{\textcolor[rgb]{0.40,0.40,1.00}{##1}}}
\@namedef{PYG@tok@o}{\def\PYG@tc##1{\textcolor[rgb]{0.20,0.20,0.20}{##1}}}
\@namedef{PYG@tok@ow}{\let\PYG@bf=\textbf\def\PYG@tc##1{\textcolor[rgb]{0.00,0.00,0.00}{##1}}}
\@namedef{PYG@tok@nb}{\def\PYG@tc##1{\textcolor[rgb]{0.00,0.47,0.13}{##1}}}
\@namedef{PYG@tok@nf}{\let\PYG@bf=\textbf\def\PYG@tc##1{\textcolor[rgb]{0.33,0.93,0.87}{##1}}}
\@namedef{PYG@tok@nc}{\let\PYG@bf=\textbf\def\PYG@tc##1{\textcolor[rgb]{0.93,0.60,0.93}{##1}}}
\@namedef{PYG@tok@nn}{\let\PYG@bf=\textbf\def\PYG@tc##1{\textcolor[rgb]{0.05,0.52,0.71}{##1}}}
\@namedef{PYG@tok@ne}{\let\PYG@bf=\textbf\def\PYG@tc##1{\textcolor[rgb]{1.00,0.00,0.00}{##1}}}
\@namedef{PYG@tok@nv}{\def\PYG@tc##1{\textcolor[rgb]{0.00,0.20,0.40}{##1}}}
\@namedef{PYG@tok@vi}{\def\PYG@tc##1{\textcolor[rgb]{0.67,0.67,1.00}{##1}}}
\@namedef{PYG@tok@vc}{\def\PYG@tc##1{\textcolor[rgb]{0.80,0.80,1.00}{##1}}}
\@namedef{PYG@tok@vg}{\def\PYG@tc##1{\textcolor[rgb]{1.00,0.53,0.27}{##1}}}
\@namedef{PYG@tok@no}{\let\PYG@bf=\textbf\def\PYG@tc##1{\textcolor[rgb]{0.33,0.93,0.87}{##1}}}
\@namedef{PYG@tok@nl}{\let\PYG@bf=\textbf\def\PYG@tc##1{\textcolor[rgb]{0.60,0.47,0.00}{##1}}}
\@namedef{PYG@tok@ni}{\def\PYG@tc##1{\textcolor[rgb]{0.53,0.00,0.00}{##1}}}
\@namedef{PYG@tok@na}{\def\PYG@tc##1{\textcolor[rgb]{0.00,0.00,0.47}{##1}}}
\@namedef{PYG@tok@nt}{\def\PYG@tc##1{\textcolor[rgb]{0.00,0.47,0.00}{##1}}}
\@namedef{PYG@tok@nd}{\let\PYG@bf=\textbf\def\PYG@tc##1{\textcolor[rgb]{0.33,0.33,0.33}{##1}}}
\@namedef{PYG@tok@s}{\def\PYG@bc##1{{\setlength{\fboxsep}{0pt}\colorbox[rgb]{0.88,0.88,1.00}{\strut ##1}}}}
\@namedef{PYG@tok@sc}{\def\PYG@tc##1{\textcolor[rgb]{0.53,0.53,1.00}{##1}}}
\@namedef{PYG@tok@sd}{\def\PYG@tc##1{\textcolor[rgb]{0.87,0.27,0.13}{##1}}}
\@namedef{PYG@tok@si}{\def\PYG@bc##1{{\setlength{\fboxsep}{0pt}\colorbox[rgb]{0.93,0.93,0.93}{\strut ##1}}}}
\@namedef{PYG@tok@se}{\let\PYG@bf=\textbf\def\PYG@tc##1{\textcolor[rgb]{0.40,0.40,0.40}{##1}}\def\PYG@bc##1{{\setlength{\fboxsep}{0pt}\colorbox[rgb]{0.88,0.88,1.00}{\strut ##1}}}}
\@namedef{PYG@tok@sr}{\def\PYG@tc##1{\textcolor[rgb]{0.00,0.00,0.00}{##1}}\def\PYG@bc##1{{\setlength{\fboxsep}{0pt}\colorbox[rgb]{0.88,0.88,1.00}{\strut ##1}}}}
\@namedef{PYG@tok@ss}{\def\PYG@tc##1{\textcolor[rgb]{1.00,0.80,0.53}{##1}}}
\@namedef{PYG@tok@sx}{\def\PYG@tc##1{\textcolor[rgb]{1.00,0.53,0.53}{##1}}\def\PYG@bc##1{{\setlength{\fboxsep}{0pt}\colorbox[rgb]{0.88,0.88,1.00}{\strut ##1}}}}
\@namedef{PYG@tok@m}{\let\PYG@bf=\textbf\def\PYG@tc##1{\textcolor[rgb]{0.40,0.00,0.93}{##1}}}
\@namedef{PYG@tok@mi}{\let\PYG@bf=\textbf\def\PYG@tc##1{\textcolor[rgb]{0.40,0.40,1.00}{##1}}}
\@namedef{PYG@tok@mf}{\let\PYG@bf=\textbf\def\PYG@tc##1{\textcolor[rgb]{0.40,0.00,0.93}{##1}}}
\@namedef{PYG@tok@mh}{\let\PYG@bf=\textbf\def\PYG@tc##1{\textcolor[rgb]{0.00,0.33,0.53}{##1}}}
\@namedef{PYG@tok@mo}{\let\PYG@bf=\textbf\def\PYG@tc##1{\textcolor[rgb]{0.27,0.00,0.93}{##1}}}
\@namedef{PYG@tok@gh}{\let\PYG@bf=\textbf\def\PYG@tc##1{\textcolor[rgb]{0.00,0.00,0.50}{##1}}}
\@namedef{PYG@tok@gu}{\let\PYG@bf=\textbf\def\PYG@tc##1{\textcolor[rgb]{0.50,0.00,0.50}{##1}}}
\@namedef{PYG@tok@gd}{\def\PYG@tc##1{\textcolor[rgb]{0.63,0.00,0.00}{##1}}}
\@namedef{PYG@tok@gi}{\def\PYG@tc##1{\textcolor[rgb]{0.00,0.63,0.00}{##1}}}
\@namedef{PYG@tok@gr}{\def\PYG@tc##1{\textcolor[rgb]{1.00,0.00,0.00}{##1}}}
\@namedef{PYG@tok@ge}{\let\PYG@it=\textit}
\@namedef{PYG@tok@gs}{\let\PYG@bf=\textbf}
\@namedef{PYG@tok@gp}{\let\PYG@bf=\textbf\def\PYG@tc##1{\textcolor[rgb]{0.78,0.36,0.04}{##1}}}
\@namedef{PYG@tok@go}{\def\PYG@tc##1{\textcolor[rgb]{0.53,0.53,0.53}{##1}}}
\@namedef{PYG@tok@gt}{\def\PYG@tc##1{\textcolor[rgb]{0.00,0.27,0.87}{##1}}}
\@namedef{PYG@tok@err}{\def\PYG@tc##1{\textcolor[rgb]{1.00,0.00,0.00}{##1}}\def\PYG@bc##1{{\setlength{\fboxsep}{0pt}\colorbox[rgb]{1.00,0.67,0.67}{\strut ##1}}}}
\@namedef{PYG@tok@kc}{\let\PYG@bf=\textbf\def\PYG@tc##1{\textcolor[rgb]{0.13,0.53,0.60}{##1}}}
\@namedef{PYG@tok@kd}{\let\PYG@bf=\textbf\def\PYG@tc##1{\textcolor[rgb]{0.13,0.53,0.60}{##1}}}
\@namedef{PYG@tok@kn}{\let\PYG@bf=\textbf\def\PYG@tc##1{\textcolor[rgb]{0.13,0.53,0.60}{##1}}}
\@namedef{PYG@tok@kr}{\let\PYG@bf=\textbf\def\PYG@tc##1{\textcolor[rgb]{0.13,0.53,0.60}{##1}}}
\@namedef{PYG@tok@bp}{\def\PYG@tc##1{\textcolor[rgb]{0.00,0.47,0.13}{##1}}}
\@namedef{PYG@tok@fm}{\let\PYG@bf=\textbf\def\PYG@tc##1{\textcolor[rgb]{0.33,0.93,0.87}{##1}}}
\@namedef{PYG@tok@vm}{\def\PYG@tc##1{\textcolor[rgb]{0.00,0.20,0.40}{##1}}}
\@namedef{PYG@tok@sa}{\def\PYG@bc##1{{\setlength{\fboxsep}{0pt}\colorbox[rgb]{0.88,0.88,1.00}{\strut ##1}}}}
\@namedef{PYG@tok@sb}{\def\PYG@bc##1{{\setlength{\fboxsep}{0pt}\colorbox[rgb]{0.88,0.88,1.00}{\strut ##1}}}}
\@namedef{PYG@tok@dl}{\def\PYG@bc##1{{\setlength{\fboxsep}{0pt}\colorbox[rgb]{0.88,0.88,1.00}{\strut ##1}}}}
\@namedef{PYG@tok@s2}{\def\PYG@bc##1{{\setlength{\fboxsep}{0pt}\colorbox[rgb]{0.88,0.88,1.00}{\strut ##1}}}}
\@namedef{PYG@tok@sh}{\def\PYG@bc##1{{\setlength{\fboxsep}{0pt}\colorbox[rgb]{0.88,0.88,1.00}{\strut ##1}}}}
\@namedef{PYG@tok@s1}{\def\PYG@bc##1{{\setlength{\fboxsep}{0pt}\colorbox[rgb]{0.88,0.88,1.00}{\strut ##1}}}}
\@namedef{PYG@tok@mb}{\let\PYG@bf=\textbf\def\PYG@tc##1{\textcolor[rgb]{0.40,0.00,0.93}{##1}}}
\@namedef{PYG@tok@il}{\let\PYG@bf=\textbf\def\PYG@tc##1{\textcolor[rgb]{0.40,0.40,1.00}{##1}}}
\@namedef{PYG@tok@ch}{\let\PYG@it=\textit\def\PYG@tc##1{\textcolor[rgb]{0.40,0.40,0.40}{##1}}}
\@namedef{PYG@tok@cm}{\let\PYG@it=\textit\def\PYG@tc##1{\textcolor[rgb]{0.40,0.40,0.40}{##1}}}
\@namedef{PYG@tok@cpf}{\let\PYG@it=\textit\def\PYG@tc##1{\textcolor[rgb]{0.40,0.40,0.40}{##1}}}
\@namedef{PYG@tok@c1}{\let\PYG@it=\textit\def\PYG@tc##1{\textcolor[rgb]{0.40,0.40,0.40}{##1}}}

\def\PYGZob{\char`\{}
\def\PYGZcb{\char`\}}

\def\PYGZhy{\char`\-}

\def\PYGZdq{\char`\"}

\makeatother

\begin{document}
	\copyrightstatement
\title{Automatic Integration of BFT State-Machine Replication into IoT Systems

\thanks{This work has received financial support by the Federal Ministry of Education and Research of Germany under grant
no 01IS18068, \sorrir{}.
All software artifacts developed are open-source available at 
\protect\url{https://github.com/sorrir/}}
}
\author{\IEEEauthorblockN{Christian Berger}
	\IEEEauthorblockA{ 
		\textit{University of Passau}\\
		Passau, Germany 
	}
	\and
	\IEEEauthorblockN{Hans P. Reiser}
	\IEEEauthorblockA{
		\textit{Reykjavik University}\\
		Reykjavik, Iceland 
	}
	\and
	\IEEEauthorblockN{Franz J. Hauck}
	\IEEEauthorblockA{
		\textit{Ulm University}\\
		Ulm, Germany 
	}
	\and
	\IEEEauthorblockN{Florian Held}
	\IEEEauthorblockA{
		\textit{Ulm University}\\
		Ulm, Germany 
	}
	\and
	\IEEEauthorblockN{Jörg Domaschka}
	\IEEEauthorblockA{
		\textit{Ulm University}\\
		Ulm, Germany 
	}
}

\maketitle

\thispagestyle{empty}
\pagestyle{plain}

\begin{abstract}
Byzantine fault tolerance (BFT) can preserve the availability and integrity of IoT systems where single components may suffer from random data corruption or attacks that can expose them to malicious behavior.
While state-of-the-art BFT state-machine replication (SMR) libraries are often tailored to fit a standard request-response interaction model with dedicated  client-server roles, in our design, we employ an IoT-fit interaction model that assumes a loosly-coupled, event-driven interaction between arbitrarily wired IoT components.

In this paper, we explore the possibility of \textit{automating and streamlining the complete process of
integrating BFT SMR into a component-based IoT execution environment}. Our main goal is providing simplicity for the developer: We strive to decouple the specification of a logical application architecture from the difficulty of incorporating BFT replication mechanisms into it.
Thus, our contributions
address the automated configuration, re-wiring and deployment of IoT components, and their replicas, within a component-based, event-driven IoT platform. 
\end{abstract}


\section{Introduction}\label{intro}
Internet-of-Things (IoT) systems consist of a variety of heterogeneous components which are connected with each other. These components may generate and consume data, provide and demand services, and collaboratively compose an application that is used by machines or people. 

Over the last years, 
IoT systems not only became popular within the public at large, e.g., by smart-home installations, but they were also incorporated in some resiliency-critical infrastructures, e.g.,
of the industrial sector~\cite{simon2017critical}. A broad field of ongoing research (e.g., see \cite{terry16towards, tsigkanos2019towards, resilienceIoTRoadmap2019})  aims for building resilient IoT systems by equipping these systems with resilience mechanisms---often by adding redundancy to mask faults.

A well-conceived resilience mechanism is state-machine replication (SMR): 
Independent and connected server replicas remain functional as a group even if up to a specific number of replicas fail~\cite{schneider1990implementing}.
A fault model describes the type of considered faults. For example,
Byzantine fault tolerance (BFT) can tolerate arbitrary and malicious behavior, even collusion among faulty replicas~\cite{lamport1982byzantine}.
When employing BFT replication, an intruder cannot break the availability or integrity of a replicated component as long as the attacker is unable to compromise more replicas than the specified bound. 

Moreover, 
BFT SMR  can serve as an important resilience mechanism for IoT infrastructures~\cite{CSUR-resilience-survey}.
For instance, it can help to protect components from random data corruption as often devices such as the Raspberry Pi are not equipped with error correcting memory. Furthermore, BFT can even mask malicious behavior which is in particularly relevant if devices are likely to be deployed in dispersed (and potentially untrustworthy) administrative domains.

Although researchers have built their own BFT SMR frameworks, there are only few available for public use, e.g.,~\cite{bessani2014state,barger2021smartbft, aksoy2019aegean, stathakopoulou2019mir}, and they are not easy to use.
In fact, these frameworks do not specifically target the requirements of IoT applications due to the following reasons:

(1) A firm separation of \textit{client-server} roles  is assumed instead of considering an architecture of individual IoT components that can be arbitrarily connected (and due to the high dynamics of IoT, this architecture may easily evolve over time thus requiring flexible adaptations).

(2) The interaction model is often assumed to be a \textit{request-response} pattern instead of a loosely-coupled, event-driven interaction model, which is typically used in the IoT domain.

(3) Replication has to be considered from the beginning \textit{during development} and cannot be configured later as an orthogonal option, although the latter would hide the burden of integrating desired replication mechanisms into the software. 

(4) The deployment of replicas is completely disregarded---yet it is important in the IoT to be capable of mapping a desired resilient software architecture to the available hardware landscape, where devices may impose certain constraints towards the deployment process making it non-trivial.

Overall, we conclude that ``classical'' SMR frameworks are no perfect fit for rapid development of IoT applications due to these shortcomings. We propose that \textit{development can be both simplified and accelerated} by automating the relevant steps necessary to integrate BFT requirements and mechanics, e.g., library configuration, replica management, key distribution, input dissemination, output consolidation and deployment decisions (distribution of replicas to available hardware devices).

By employing the commonly used 
SMR library \bftsmart~\cite{bessani2014state} as building block, we present a method that simplifies and automates the integration of BFT SMR in a component-based, event-driven IoT framework, thus alleviating the incorporation of resilience in IoT applications.

Further, we provide a concrete, practical implementation of our method, that is publicly available for the open-source \sorrir{} framework\cite{domaschka2019sorrir}, which particularly serves as a model of a component-based and event-driven IoT framework.

\subsection*{Contributions and Structure of this Paper}

Our approach addresses the mentioned challenges by starting with a \textit{logical software architecture} (\textsc{LSA}) describing application components and their connections.
As a second and independent step, an \textsc{LSA} is transformed into a \textit{replication-enriched software architecture} (\textsc{ReSA}), enhancing the \textsc{LSA} by replication mechanisms, additional components and their configurations, finally mapped to the hardware landscape of available devices.
Our approach

\begin{itemize}
    \item proposes a building-block principle for state-machine replication that dissects replication into its functional parts which can be flexibly glued to IoT components;
    \item seamlessly integrates customizable BFT replication, based on the replication library \bftsmart{}, into a component-based event-driven IoT architecture; 
    \item automatically deploys component replicas on the hardware architecture, recognizing the constraints of the available IoT devices.
\end{itemize}






At first, we provide related work in Section~\ref{related-work} and  relevant background knowledge helpful to follow our approach as well as our system model in Section~\ref{background}. Subsequently, we present all technical details of our approach in Section~\ref{solution}, and later evaluate our implemented solution in Section~\ref{evaluation}. 
Finally, we draw our conclusions in Section~\ref{conclusion}.

\section{Related Work} \label{related-work}
Existing BFT SMR frameworks~\cite{bessani2014state,barger2021smartbft, aksoy2019aegean, stathakopoulou2019mir} can not directly be used for rapid development of component-based, event-driven IoT applications.
There are a few research works that touch the questions of either making SMR more viable for component-based architectures, or   integrating replication mechanisms (though not necessarily BFT) into IoT environments.

An early system providing a platform for replicated software components is \emph{Delta-4}~\cite{barret1990delta4}.
A modern approach is to use containerised service replicas and manage them by \emph{Kubernetes}~\cite{netto2017smrkubernetes}.
Both approaches support a very vague definition of software component, whereas \sorrir{} has a well-defined component model that is here transparently enhanced by replication.
Within the component-based cloud platform \emph{COSCA}~\cite{kaechele2011cosca} active replication of components was supported by special communication measures~\cite{kaechele2015coscaft}.
Derecho~\cite{jha2019derecho} is a software library for creating structured cloud services that can be replicated using the SMR approach, but cannot cope with Byzantine failures.

CEFIoT~\cite{javed2018cefiot} explains a crash fault-tolerant data-as-a-stream architecture that allows placing processing components
in both edge and cloud. It uses Apache Kafka for publish-subscribe messaging, so data streams can be buffered and easily replicated across a cluster.
Furthermore, SMaRt-SCADA~\cite{nogueira2018challenges} implements BFT SMR into a SCADA system, which is also an event-driven system with publish-subscribe architecture. Further, a  publish-subscribe interaction model for BFT SMR has also been studied for web applications~\cite{berger2018webbft}.


To the best of our knowledge, our approach is the first effort to streamline and automate the whole  process of integrating  BFT SMR mechanisms in component-based IoT architectures.


\section{Background and Prelimaries} \label{background}
We first explain our underlying system model~(Section~\ref{system-model}), review some relevant background on state-machine replication, in particular \bftsmart{}~(Section~\ref{smr}), which is the replication library we are using, and subsequently summarize both architecture and core characteristics of the \sorrir{} IoT framework (Section~\ref{sorrir}).

\subsection{System Model} \label{system-model}
\subsubsection*{Fault Model} Our method allows for a \textit{tunable} fault model. On the granularity level of single components, the system designer can specify whether crash faults or Byzantine faults are assumed. The Byzantine fault model includes both crash faults as well as accidental (e.g., data corruption due to memory errors) or malicious (e.g., intrusions, tampering with devices) non-crash faults and thus may be a preferable choice when component replicas are deployed in dispersed administrative domains. Tolerating $f$ faulty replicas needs a replica group of $n$ replicas composing the replicated component with $n\geq{}3f+1$ (BFT) or $n\geq{}2f+1$ (CFT) respectively~\cite{bessani2014state}, and the system designer can choose $f$ as a trade-off between fault tolerance and the replication overhead.

\subsubsection*{Network Model} We assume the partially synchronous system model~\cite{dwork1988consensus}, in which the system can behave initially asynchronous, but after some unknown time period, called \textit{global stabilization time (GST)}, eventually some upper bound~$\delta$ on communication delay holds. As soon as \textit{GST} is reached the used replication library guarantees that progress can be made, i.e., messages can be delivered in total order. The communication is point-to-point, authenticated and reliable.

\subsection{State-Machine Replication} \label{smr}
State-machine replication is based on independent servers hosting replicas of the same service\cite{schneider1990implementing}.
Client requests are delivered to all correct replicas in identical (total) order, so that their deterministic execution in each replica leads to equal replica states.
There are multicast protocols that are able to support CFT and BFT.
In case of BFT, the client needs to receive $f+1$ identical answers from different replicas.

\bftsmart{} is a framework library to develop SMR-based applications in Java\cite{bessani2014state}.
It can be used for CFT and BFT, and provides appropriate consensus-based multicast protocols.
In particular, we employ the decision forwarding extension~\cite{berger2021making}.
Communication is authenticated and encrypted by using TLS, where each replica possesses an RSA key pair ($sk_i$, $pk_i$) to construct secure channels with all the other replicas.

\subsection{The \sorrir{} Framework} \label{sorrir}

\begin{figure}[t]
	\centering
	\includegraphics[width=1\linewidth]{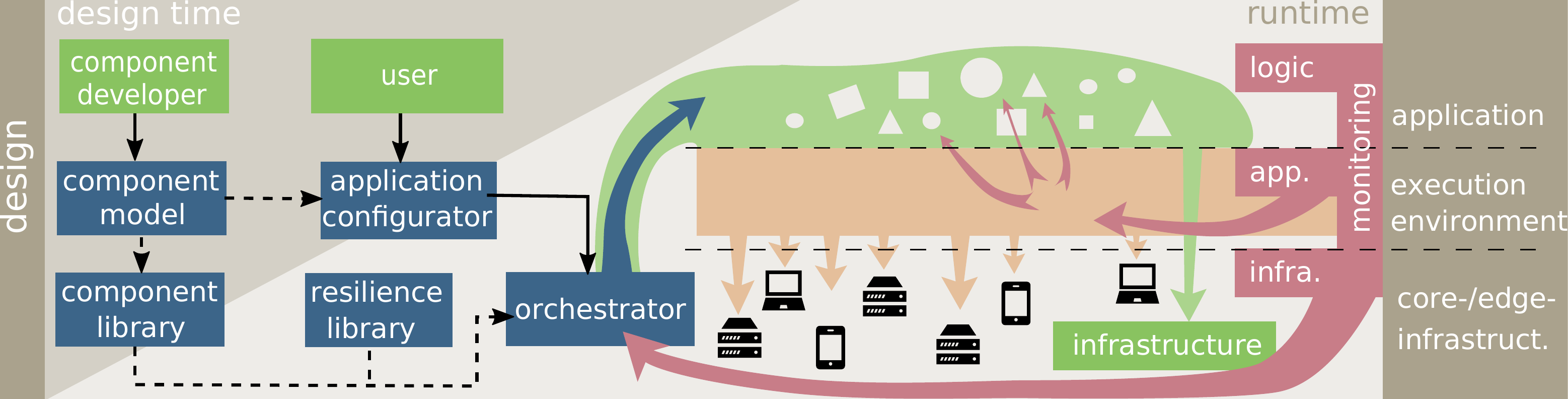}
	\caption{High-level overview of the \sorrir{} approach~\cite{domaschka2019sorrir}.}
	\label{fig:sorrir}
\end{figure}

\sorrir{} is a research framework for developing resilient IoT applications\cite{domaschka2019sorrir}.
Applications are composed from software components that interact by sending and receiving events\cite{tichy2020experiences}.
Components may be organised in a hierarchy so that some can be grouped to form a composite component that can be used like any other component. Components define ports over which they can be wired together by connections, following a building-block concept.
Each component can be attached to resilience mechanisms from a resilience library.
The main idea is to keep application code as independent from resilience mechanisms as possible.
Developers can build applications from components with an application configurator (\textsc{LSA}), and attach resilience mechanisms to each component, leading to a \textsc{ReSA}.
At runtime, an orchestrator takes care that components and resilience mechanisms are deployed on the available infrastructure of an IoT application. 
A symbolic overview can be found in Fig.~\ref{fig:sorrir}.

Additionally, \sorrir{} has a component-dependent degradation concept.
For the case of failures, developers can specify levels of degraded behaviour. 
\sorrir{} then automatically switches components into degradation levels, if certain conditions occur, e.g., dependent components fail or degrade\cite{hess2021morpheus}.

In \sorrir{}, each component is modeled as a state machine. 
In the execution engine underlying the \sorrir{} framework 

(1)~events are executed step-wise in a loop  until no longer a state transition can be applied and all components deployed on the \sorrir{} unit have reached their stable state, 

(2)~the execution of any incoming event (that arrived on any arbitrary port) needs to complete before the execution of the next incoming event is allowed to start, and 

(3)~the execution of any event itself is deterministic which is a strict requirement for all components.

The \sorrir{} framework is implemented in TypeScript on Node.js and provides different protocols for transmitting events between remote IoT components, i.e,. HTTP, MQTT and BLE.
Components are placed into so called \emph{units} that are finally implemented as containers.
Thus, units are the smallest distributable entities and can contain one or more components.

\section{The SORRIR Approach to SMR} \label{solution}

	Our approach for incorporating replication mechanisms for desired components consists of a streamlined process of solving individual challenges: First, we utilize a building-block principle (Section~\ref{architecture}) that allows us to modularly swap-in \textit{component replicas} within the logical software architecture (\textsc{LSA}). Second, we employ a resilience configuration (Section~\ref{resilience_configuration}) that lets developers customize replication mechanisms for individual components. Third, we subsequently conduct an automated integration procedure of the specified replication mechanisms by transforming the \textsc{LSA} into a replication-enriched software architecture (\textsc{ReSA}) that contains instances of component replicas and dedicated output-consolidating components as well as  instances of the \bftsmart{} middleware (Section~\ref{integration}). In this step, components are also automatically re-wired to account for swapping in the newly generated component replicas and consolidators.
	Fourth, we use the k3s\footnote{https://k3s.io/} Kubernetes distribution to manage the deployment of \textit{units} (instances of \sorrir{} that execute components) on available hardware devices (Section~\ref{deployment}).
	
\begin{figure}[t]
    \centering
 \hskip -1.5cm   \begin{subfigure}[b]{.45\columnwidth}
    \centering
    \includegraphics[height=1\columnwidth]{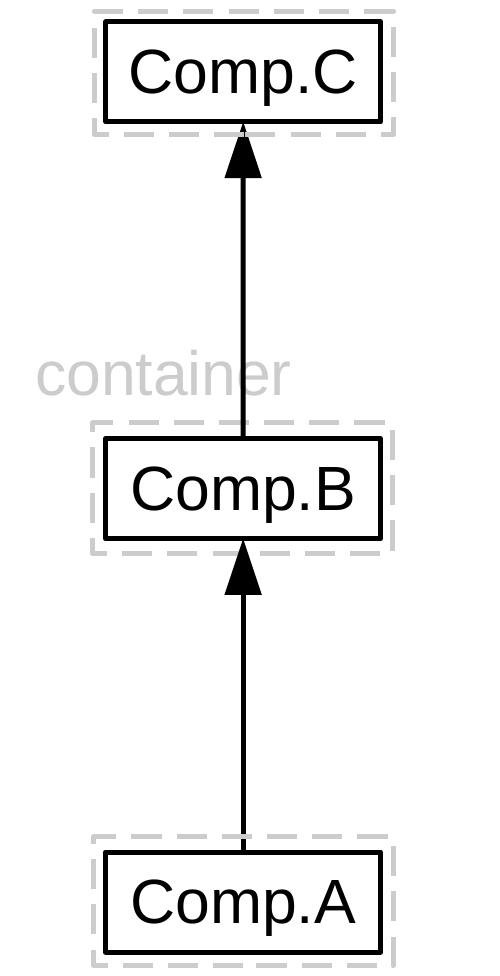}
    \caption{Logical SA.}
    \label{fig:lsa}
    \end{subfigure}
   \begin{subfigure}[b]{.45\columnwidth}
    \centering
     \includegraphics[height=1\columnwidth]{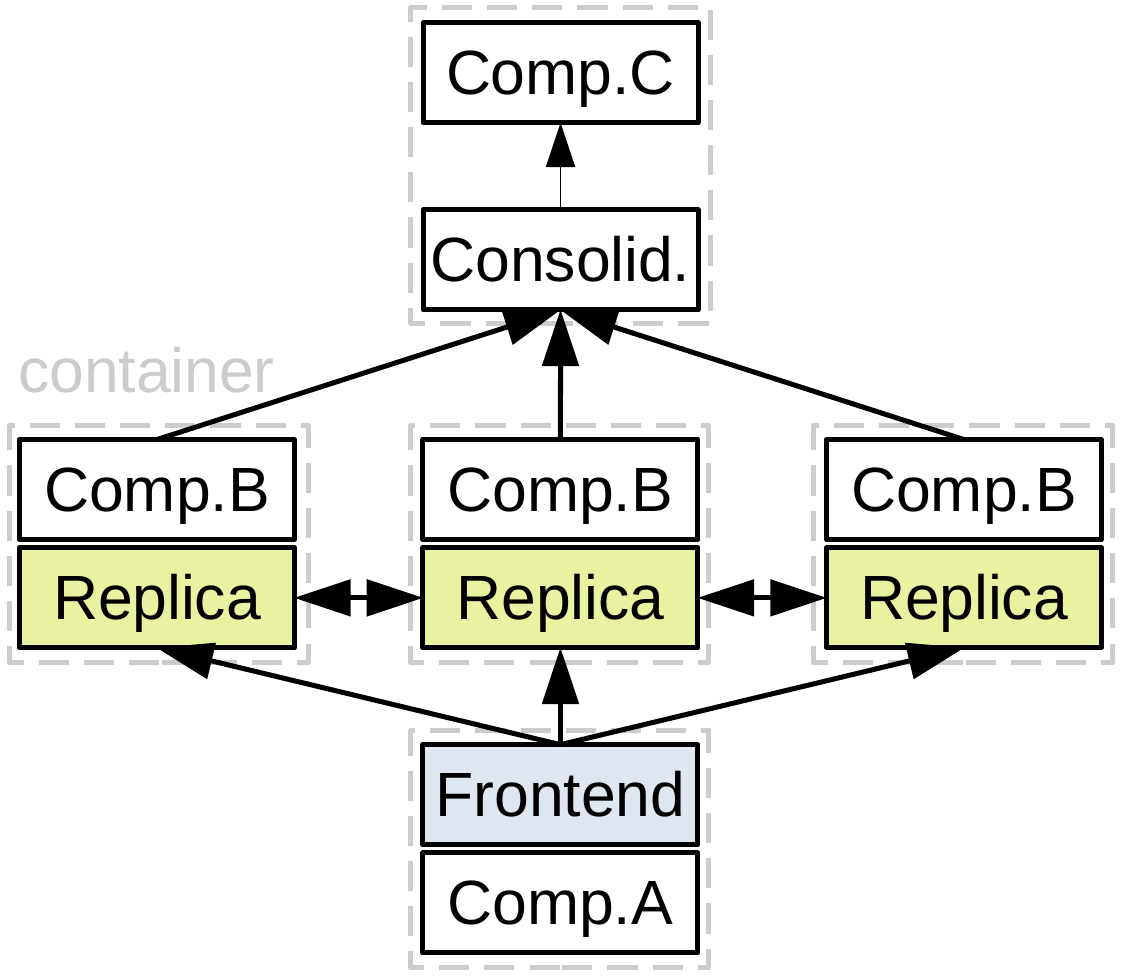}
    \caption{Replication-Enriched SA. 
    } 
    \label{fig:resa}
    \end{subfigure}
    \caption{Component-Defined Software Architecture (SA).}
    \label{fig:lsa-resa}
\end{figure}

	\begin{figure*}[hbt]
		\centering
		\includegraphics[width=0.972\linewidth]{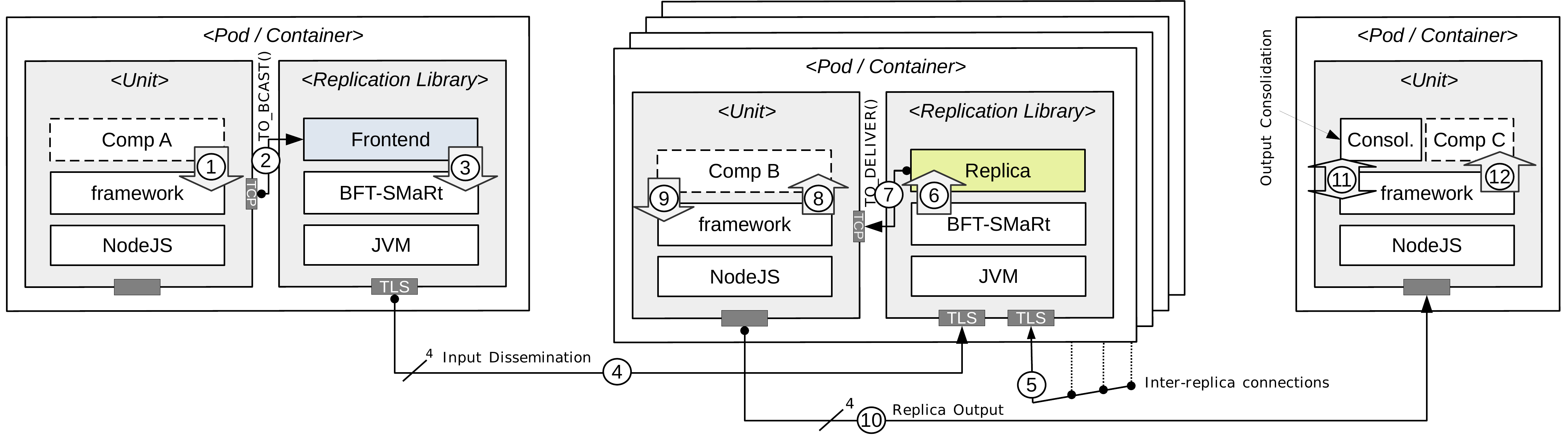}
		\caption{
A modular architecture for interaction between SMR building blocks that are instantiated on demand.}
		\label{fig:architecture}
	\end{figure*}

	\subsection{A Building-Block Principle for State-Machine Replication}
	\label{architecture}
	  The core idea of our architecture is making the existing BFT replication library \bftsmart~\cite{bessani2014state} viable for practical IoT applications while satisfying the goals listed before.

	  Our solution employs modularity by utilizing a building-block principle in which different building blocks implement BFT replication mechanics.
	  There is the \textsc{Frontend} (used for input dissemination), the \textsc{Replica} (framework instance of \bftsmart{} used as a proxy to the actual component) and the \textsc{Consolidator} (used for output consolidation).
	  These components are inserted \textit{on demand} into the architecture.
	  Subsequently, connections between components are re-wired to account for both newly inserted and replicated components.
	  To explain the interplay of these building blocks as well as their interaction with the \sorrir{} framework and the \bftsmart{} library, we use in the following a simple example of three components \textsc{Comp~A}, \textsc{Comp~B}, and \textsc{Comp~C} sending events from A to B and from B to C (Fig.~\ref{fig:lsa}).
	  In the corresponding \textsc{ReSA}, \textsc{Comp~B} is replicated (Fig.~\ref{fig:resa}).
	  Fig.~\ref{fig:architecture} shows the technical details of this configuration.

	 		 \textsc{Component}: Application components (\tikz \filldraw [fill=white, draw=black, dashed] (0.1,0.1) rectangle (0.3,0.3); in Fig.~\ref{fig:architecture}) are usually created by the developer. A necessary requirement is that they provide a deterministic \texttt{step()} function which is used to transition the component state (see \sorrir{} programming model~\cite{tichy2020experiences}). No additional interfaces are required or needed.
	 		The communication and execution of components is handled by the underlying \sorrir{} framework \circled{1} that offers a variety of different communication technologies. Since \textsc{Comp~A} sends an event to the replicated \textsc{Comp~B}, the \sorrir{} framework delegates this event to an instance of \textsc{Frontend}~\circled{2}, which disseminates it as input to all replicas.
	 		
	 		\textsc{Frontend}:
	 		The \sorrir{} framework treats the \textsc{Frontend} like a \textit{communication technology}  over which it can send events.
	 		Specifically, the \textsc{Frontend} acts as a local proxy when an application component is speaking to a replicated component, thus broadcasting the event to all replicas. 
	 		Thus, it hides the fact that the event is sent to a replicated component.
	 		For the \sorrir{} framework, the forwarding of an event to a \textsc{Frontend} looks like ordinary event communication.
	 		The \textsc{Frontend} calls the \texttt{invokeOrdered()} method of the client-side \bftsmart{} library \circled{3} so that the event passes the \textit{total-order multicast} layer. 
	 		Each \textsc{Replica} instance of the replicated component \textsc{Comp~B} receives the event~\circled{4}.
	 		
	 		

	 		\textsc{Replica}:
	  The \textsc{Replica} implements the server-side of the \bftsmart{} protocol. It receives messages from frontends and communicates with the other replicas in a three phase agreement pattern to decide a consensus instance. The \bftsmart{} leader drains messages from its in-queues and puts them in a batch, which the leader subsequently proposes to all other replicas, thus determining the order of execution \circled{5}. 
	  
	  Once a consensus is decided, messages contained in the batch are delivered in order to the  \texttt{executeOrdered()} method of \textsc{BFT-SMaRt} \circled{6}. The \textsc{Replica} implementation forwards ordered events back to the \sorrir{} framework \circled{7}, which subsequently pushes them one after another into the event queue of the receiving component \circled{8} (e.g., a replica of the application component \textsc{Comp~B}). 
	  In our example, all replicas of \textsc{Comp~B} process the same event, and produce some outgoing event to be sent to \textsc{Comp~C}, which is handled by the \sorrir{} communication system \circled{9}. This could result in a HTTPS POST request~\circled{10}, and the same event is received multiple times by the framework of the receiver \textsc{Comp~C}.
	 		
	 		 
	 		\textsc{Consolidator}: The \textsc{Consolidator} is a \sorrir{}-provided component that is automatically instantiated for any component that \textit{receives} from a replicated component. Its purpose is to provide the abstraction 
	 		of receiving events as if a non-replicated sender was involved.
	 		It consolidates the output of several replicas into a single input towards the receiving application component  (here: \textsc{Comp~C}). On a technical level, the consolidator checks for a certain threshold (depending on the configured fault model) of \textit{matching} events \circled{11} (e.g., same sequence number, same content), de-duplicates them, and subsequently puts the \textit{consolidated event} into the event queue of the intended receiver---in our example \textsc{Comp~C} \circled{12}.
	 		
	 		We provide two implementations, 
	 		used for the BFT or CFT model, respectively. The BFT consolidator employs a strict comparison of an event's content, thus masking deviating output of faulty components. 
	 		This modular approach allows developers to inject their own consolidator component to realize application-specific behavior such as consolidating non-deterministic input. This is useful, e.g., when reading from a replicated sensor and checking if obtained values are within a certain interval (thus relaxing the strict consistency of outputs as they are expected in traditional SMR). The communication and execution of a consolidator component is handled by the underlying \sorrir{} framework.

	\sorrir{} framework:
	The \sorrir{} framework is the middleware in which all other building blocks are plugged into. It serves as an execution platform that steps through components, executing events until no further state transition is possible. Also, it handles the exchange of messages with other components, either deployed locally or located on a remote unit. It comes with an configurator that allows developers to wire components together, thus accelerating development.

\subsection{Configuration of the Replication Mechanisms}
\label{resilience_configuration}
A \textit{resilience configuration} is a simple specification of which resilience mechanisms should be applied to which components.
This configuration is thus a declarative description how to transform an \textsc{LSA} to a \textsc{ReSA} (cf. Fig.~\ref{fig:lsa-resa}).
Note, that we make SMR configurable on the granularity of individual components. This means we can easily employ different fault models, replica group sizes or consolidation techniques for different components within our software architecture.

The mechanism-specific properties specify the intended behavior, e.g., number $f$ of faulty component-replicas to be tolerated, the employed fault model (Byzantine or Crash) and as an option an alternative consolidator component.
In Figure~\ref{fig:resilience_config}, we show the example of a simple configuration which specifies that component \textsc{Comp-B} is being replicated to tolerate a single faulty component replica within the Byzantine fault model.

\begin{figure}[h]
    \centering
\begin{Verbatim}[commandchars=\\\{\}] 
\PYG{p}{\PYGZob{}}\PYG{n+nt}{\PYGZdq{}components\PYGZdq{}}\PYG{p}{:}\PYG{+w}{ }\PYG{p}{[}
\PYG{+w}{  }\PYG{p}{\PYGZob{}}\PYG{n+nt}{\PYGZdq{}id\PYGZdq{}}\PYG{p}{:}\PYG{+w}{ }\PYG{l+s+s2}{\PYGZdq{}Comp\PYGZhy{}B\PYGZdq{}}\PYG{p}{,}
\PYG{+w}{   }\PYG{n+nt}{\PYGZdq{}mechanisms\PYGZdq{}}\PYG{p}{:}\PYG{+w}{ }\PYG{p}{\PYGZob{}}
\PYG{+w}{     }\PYG{n+nt}{\PYGZdq{}activeReplication\PYGZdq{}}\PYG{p}{:}\PYG{+w}{ }\PYG{p}{\PYGZob{}}
\PYG{+w}{       }\PYG{n+nt}{\PYGZdq{}enabled\PYGZdq{}}\PYG{p}{:}\PYG{+w}{ }\PYG{k+kc}{true}\PYG{p}{,}
\PYG{+w}{       }\PYG{n+nt}{\PYGZdq{}f\PYGZdq{}}\PYG{p}{:}\PYG{+w}{ }\PYG{l+m+mi}{1}\PYG{p}{,}
\PYG{+w}{       }\PYG{n+nt}{\PYGZdq{}faultModel\PYGZdq{}}\PYG{p}{:}\PYG{+w}{ }\PYG{l+s+s2}{\PYGZdq{}BFT\PYGZdq{}}\PYG{p}{,}\PYG{+w}{	}
\PYG{+w}{       }\PYG{n+nt}{\PYGZdq{}consolidator\PYGZdq{}}\PYG{p}{:}\PYG{+w}{  }\PYG{l+s+s2}{\PYGZdq{}BFTConsolidator\PYGZdq{}}\PYG{+w}{ }\PYG{p}{\PYGZcb{}}
\PYG{+w}{ }\PYG{p}{\PYGZcb{}}\PYG{+w}{ }\PYG{p}{\PYGZcb{},}\PYG{+w}{  }\PYG{err}{...}\PYG{+w}{ }\PYG{p}{]\PYGZcb{}}
\end{Verbatim}
    \caption{The replication mechanisms can be configured on the granularity of components in a \textit{resilience configuration} file.}
    \label{fig:resilience_config}
\end{figure}

\subsection{Automatic Integration of Replication Mechanisms}
\label{integration}

Automatically integrating replication mechanisms within a component-based, event-driven IoT framework demands to provide glue code for the framework to setup the additional building blocks (see Section~\ref{architecture}) and rewiring components accordingly. 
We call this process \textsc{setup-replication}, the \textit{transformation} of the \textsc{LSA} into the \textsc{ReSA}. 
This transformation happens step-wise by iterating over the list of components, checking if a replication mechanism is specified, inserting necessary building blocks, and rewiring connections. 
 


\subsubsection{Insertion of Necessary Building Blocks}
\label{instantiateBuildingBlocks}
The transformation starts with the setup of additional building blocks on a unit. The differentiation which building blocks need to be inserted into some unit $u$ follows a set of rules:
\begin{itemize}
	\item \textsc{Consolidator}: Insert for each component on $u$ that receives from a replicated component (to be more concrete: for each replicated component it receives from). 
	\item \textsc{Frontend}: Insert for each component on $u$ that sends to a replicated component (in particular: for each replicated component it sends to).
	\item \textsc{Replica}: Insert iff a component replica instance of some replicated component is to be deployed on $u$ (this depends on deployment decisions, which we will explain in more detail in Section \ref{deployment}). 
\end{itemize}

In \sorrir{} the \textsc{LSA} specifies for each unit a locally deployed configuration \textsc{conf} which consists of components and their connections.
This \textsc{conf} is expanded by our provided \textsc{Consolidator} components and component replicas.
Also, \textsc{setup-replication} creates \bftsmart{} configuration files and bootstraps necessary \textsc{Frontend}s and \textsc{Replica}s.

After the setup procedure, the configuration still contains the old connections and does not yet consider the newly inserted building blocks. To complete the transformation to the \textsc{ReSA}, rewiring of the components is necessary.

\subsubsection{Rewiring Connections}
\label{rewireConnections}
In the next step, the overall architecture of every unit is being rewired to satisfy two goals: (1)~input dissemination and (2)~output consolidation. In particular, this concerns the following connections between \textit{components} at each \textit{unit}:

\paragraph{Input dissemination} If component $c$ has an outgoing connection to some replicated component $r$ then $c$'s connection is re-configured to employ the \textit{total-order multicast} communication technology. This means an event is disseminated to
 \textit{all} ($n_r$) component replica instances using the \textsc{Frontend} $F_{c,r}$. Further, if component $c$ is a replicated component, then all of its incoming connections 
 are configured to employ \textit{total-order delivery} of events over the local \textsc{Replica} $R_c$ instance that is connected to the \sorrir{} framework. 
	
\paragraph{Output consolidation} If component $c$ has an incoming connection from replicated component $r$ then $c$ is being re-wired to receive from the newly created consolidator component $\textit{Cons}_{r,c}$ instead. Further, component $\textit{Cons}_{r,c}$ is being wired to receive from 
 \textit{all} component replicas of $r$ which creates $n_r$ additional connections.

\begin{figure}[h]
    \centering
    \includegraphics[width=0.9\columnwidth]{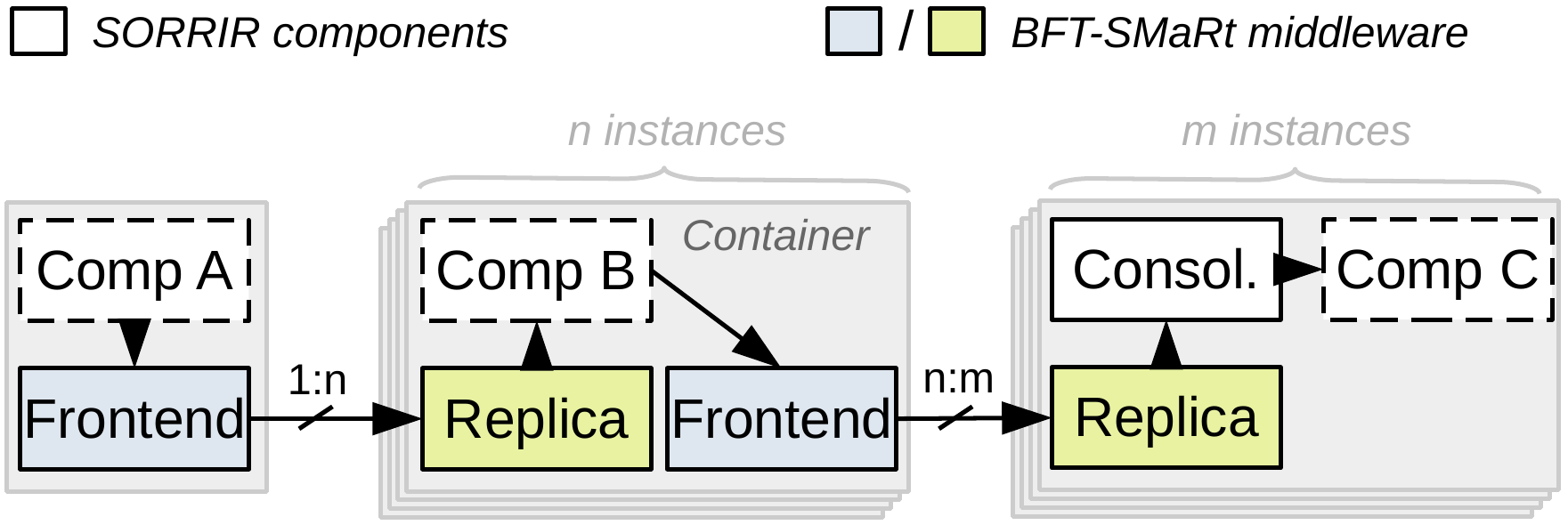}
    \caption{Interaction between two replication groups.}
    \label{fig:group_interaction}
\end{figure}

\paragraph{Interaction between two replicated components}
 We support the interaction between two replicated components. To illustrate the behavior (see Fig.~\ref{fig:group_interaction}), let us suppose that in an application a replicated component, defined as a set of state machines $\textsc{B} = \{b_1, .., b_n\}$, is wired over a directed connection to another replicated component $\textsc{C}= \{c_1,..,c_m\}$. We denote this connection by $ \textsc{B} \rightarrow  \textsc{C}$.

For each $i$ with $1 \leq i \leq n$ and each $j$ with $1 \leq j \leq m$ a connection from the Fronted $F_{i,j}$ that 
is instantiated for $b_i$ to a Replica instance $R_j$ co-located with component $c_j$ is created. Any event $e$ that is sent through $ \textsc{B} \rightarrow  \textsc{C}$ arrives up to $n$-times at each of the $m$ Replicas. After ordering, events are passed to the \sorrir{} framework, de-duplication is performed by the consolidators and subsequently $\forall j \in \{1..m\}$ only once $e$ is delivered to each component instance $c_j$.

	
\subsection{Automated Deployment}
\label{deployment}

Our automated deployment approach is general enough for most IoT systems. The only restriction is that the system must be managed by \textit{k3s} and that associated bundled platform components (i.e. \textit{kubelet, kube-proxy}) are installed on the individual devices. \textit{k3s} is a special \textit{Kubernetes} distribution, to manage and execute containerized workloads in IoT edge environments. Today, Kubernetes is seen as the de-facto standard solution for container orchestration technologies~\cite{ferreira2019performance} and helps us to manage an IoT \textit{testbed} that is built from multiple heterogeneous devices like \textit{Jetson TX2} or \textit{Raspberry Pi 4b} single board computers. Further, k3s provides meta-information on devices like \textit{device name, type, architecture} and custom tags like \textit{location}, which can be exploited for the selection and the deployment of BFT component replicas. 

Next, we describe the workflow from configuring an application that incorporates BFT replicated state machines up to the orchestration step complying to aforementioned boundary conditions associated with BFT replication. On a conceptual level, we describe how in the orchestration step the software architecture as defined in the \textsc{ReSA} gets decoupled from the decision of component placement on individual devices.

\subsubsection{Configuration and Orchestration} We use the \sorrir{} Configurator\footnote{\url{https://github.com/sorrir/configurator}} in the form of a REST API to configure the logical software architecture of the application. Precisely, we build the application by adding components and their connections and specify 
their assignment to units.
Then, we add the BFT state-machine replication as resilience mechanism for some specific component (by referring to its identifier) to the resilience configuration (see Fig~\ref{fig:resilience_config}).
Additionally, we can set the BFT parameters for $
n$ and $f$ (see Section~\ref{system-model}).

In the next step, the \sorrir{} \textit{Orchestrator}\footnote{\url{https://github.com/sorrir/orchestrator}} uses this configuration to create specific resources in the k3s ecosystem. These are \textit{deployments} that
create \textit{pods} in which the units get encapsulated. As k3s uses containerized software components, the units get deployed in the form of containers (see Fig.~\ref{fig:architecture}).

Kubernetes \textit{Services} act as a frontend to individual pods, respectively units, in the sense that they route incoming messages to the individual pods. In the \sorrir{} case, we always map one service to one pod (i.e. unit).
Eventually, resources named \textit{configMaps} are created alongside.
These serve as an abstraction to inject configuration files for our \sorrir{} application inside the containers. 

\subsubsection{Deployment of BFT-associated Components}
    The orchestrator deploys all different entities as shown in Fig.~\ref{fig:architecture} encapsulated in \textit{pods} on different devices. 
    In particular, the orchestrator takes care that different software parts for BFT replication, e.g., \textsc{Frontend}, \textsc{Replica} and \textsc{Consolidator} are correctly utilized and wired together.
    Apart of the connection inside a container, that is between a \textit{\texttt{<}Unit\texttt{>}} and the
    \textit{\texttt{<}Replication Library\texttt{>}} via a local socket, the orchestrator exploits built-in k3s functionality for setting up a proper IP-based communication system. This is done through an integrated DNS server and the
    created \textit{services}, which are bound to individual pods. The \textit{services} can be addressed by means of DNS resolvable names that are identical to the associated pod's unit name with the main task to forward received messages to its assigned pods. The service names are used within the \bftsmart{} host configuration file which is deployed by the orchestrator on all components that are part of a specific replication group (either as invoker or replica).
    This ensures that component replicas can establish connections to each other by utilizing their service names, as well as the sender component can establish a connection to every component replica.
    The \textit{deployments} are responsible for setting the container image URI and, if needed, for pointing to a \textit{Kubernetes secret} that encrypts user credentials for accessing the container registry. Finally, \textit{configMaps} enable the creation of configuration files, that get injected into the containers. For example, the component instance to be replicated and its parameters $n$ and $f$ 
    are specified by this.
    
    The orchestrator takes over control of the key generation and distribution. Since the authenticated and encrypted communication between involved component replicas as described in Section~\ref{smr} uses symmetric secrets which are negotiated through \textit{Signed Diffie-Hellman}, we need the a-priori existence of asymmetric key material: 
    The orchestrator generates an RSA key pair ($sk_i$, $pk_i$) for each component replica and deploys the private key $sk_i$ only on the respective container $i$ while all public keys are distributed to all containers that are part of the replication group.
    
    Further, 
    the orchestrator assures that replicas of the same component must not be placed on the same host. Violating this restriction, which is orthogonal to defined restrictions in the \textsc{ReSA}, would break the assumption of independent faults.


\begin{figure*}[t]
    \centering
      \begin{subfigure}[b]{.28\textwidth}
    \centering
    \input{diagrams/ordering}
    \caption{Ordering performance.}
    \label{fig:ordering}
    \end{subfigure}
    \hskip 0.2cm
    \begin{subfigure}[b]{.305\textwidth}
    \centering
    \input{diagrams/latency}
    \caption{Latency comparison.}
    \label{fig:latency}
    \end{subfigure}
       \begin{subfigure}[b]{.305\textwidth}
    \centering
    \input{diagrams/throughput}
    \caption{Throughput comparison. 
    } 
    \label{fig:throughput}
    \end{subfigure}
    \caption{Evaluation results of conducted benchmarks.}
    \label{fig:eval-results}
\end{figure*}
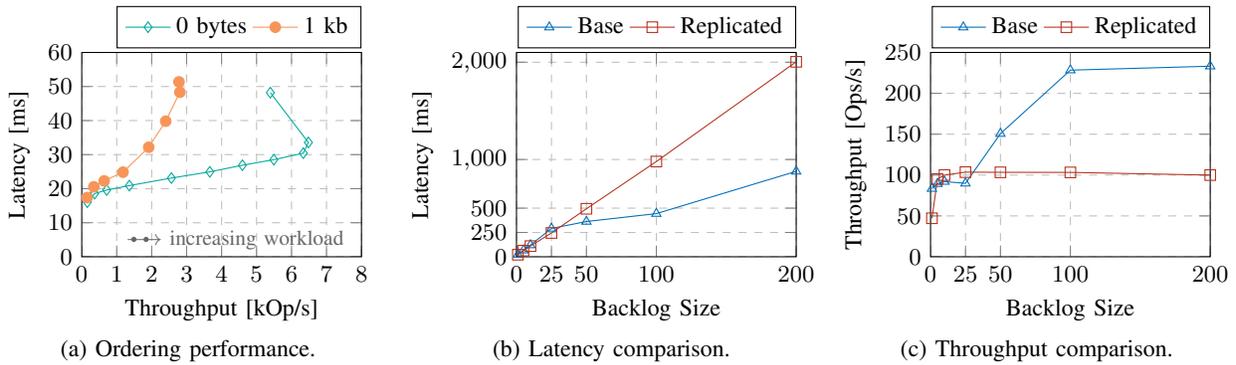

\begin{figure}[h]
    \centering
     \begin{subfigure}[b]{.22\textwidth}
    \centering
    \input{diagrams/faulty_leader_throughput}
    \caption{Throughput.}
    \label{fig:faulty-leader:throughput}
\end{subfigure}
 \begin{subfigure}[b]{.22\textwidth}
    \centering
    \input{diagrams/faulty_leader_latencies}
    \caption{Latency.}
    \label{fig:faulty-leader:latency}
\end{subfigure}
        \caption{System performance under a leader failure.}
    \label{fig:leader-failure}
\end{figure}
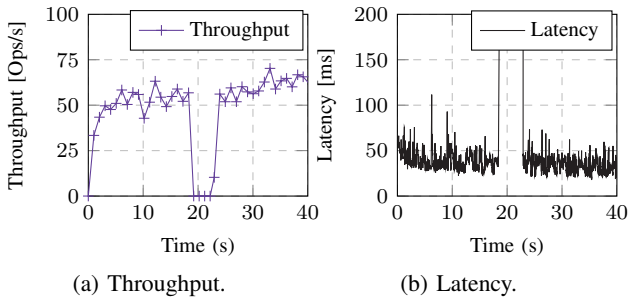

\section{Evaluation} \label{evaluation}

In this section, we first give a brief overview of our experimental setup and methods. In our first experiment we benchmark the ordering performance of the used replication library, \bftsmart{}.
Subsequently, we quantify how a replication-enriched system architecture performs compared to an architecture without replicas as a baseline to reason about the induced performance overhead. Then, we conduct an experiment in which we monitor the effects of node failure within the system. Finally, we discuss the obtained results.

\subsection{Setup and Method}
\textit{Setup.} We use a small testbed that consists of 6 Raspberry Pi 4-B single-board computers which are connected over a 100~Mbps switch. Each device is equipped with at least 4~GB RAM and is running Raspberry Pi OS 64-Bit.

\textit{Benchmarking \bftsmart.} We first conduct benchmarks for the \bftsmart{} middleware to reason about the performance of ordering events. For this purpose, we use the micro-benchmark suite 
of \bftsmart. This benchmark employs synchronous clients with request-response interaction.

A \textit{request} is a wrapper for an event (which acts as payload). 
Later, in Sections~\ref{eval:replication-overhead} and \ref{eval:leader-failure}, we only send requests (carrying events) one-way, towards the replica without caring about replies. To estimate the performance boundaries of ordering requests,  we employ request payload sizes of 0 bytes and 1~kb. 
We deploy 4 replicas on 4 different Pis and launch an increasing number of clients on the other 2 devices to stimulate the system with higher workloads. 
	
\textit{Benchmarking \sorrir.} We developed a benchmarking application in the \sorrir{} programming model, which consists of three components: 
A \textit{loadgenerator} maintains a fixed-size backlog of events that are sent simultaneously to the \textit{processor}. Increasing the size of this backlog leads to higher workloads in the system. The processor is an empty implementation that only forwards all received events to the \textit{reporter}. 
Moreover, the reporter measures and logs the system throughput and sends events back as feedback (used for controlling the backlog) to the loadgenerator: Essentially, for each event that is fed back, a new event is created. Using the feedback, the loadgenerator can calculate and log the end-to-end latency of events. 

\textit{Benchmarking a ReSA in \sorrir.}
To evaluate the ReSA, we replicate the processor in $f=1$ BFT mode, with component replicas being deployed on four separate Pis. In this setup, a consolidator is deployed on the same unit as the reporter to receive and consolidate the output from the four
 replicas, which adds additional work on the reporter unit (all handled by a single Node.js thread). An event of the benchmarking application has roughly the size of 150~bytes (which acts as payload for the underlying \bftsmart{} middleware).


\subsection{Performance of Ordering Events} \label{sect:eval:ordering}
In our first experiment we want to evaluate how the underlying replication middleware, \bftsmart{}, that is used to order events, performs on the Raspberry Pi 4-B devices. The main purpose is to understand whether the process of ordering events (which is necessary for replication) may  induce a bottleneck in systems that make use of replication.

\textit{Measurements.} We systematically increase the workload by increasing the number of synchronous clients (each client sends $5000$ requests to the replicas) until we reach the saturation point. Throughput is measured at the leader replica, while latency is measured on the client side at a chosen client. 

\textit{Results.}
We show our results in Figure~\ref{fig:ordering}.
We observe a peak throughput of $6,482$ operations per second (op/s) at a latency of $34$ ms for $0$ bytes requests and still up to $2,806$ op/s at a latency of $48$ ms for requests with 1 kb payload. These numbers indicate that \bftsmart{} replicas are able to handle the ordering of events in a sufficiently efficient way even on constrained ARM devices like the Pi 4-B.

\subsection{Replication Overhead} \label{eval:replication-overhead}
In our second experiment we compare how a ReSA performs compared to a standard non-replicated system as baseline. This is to investigate if replication mechanisms may introduce a performance drop in the system.

\textit{Measurements.} We systematically increase the size of the backlog to generate higher workloads in the system until we reach the saturation point. For this purpose we 
use 
increasing backlog sizes for our measurements.
For each backlog size, 
we configure the loadgenerator to send a total of $k=10,000$ events and use all events in the interval of [2,500, 7,500] to calculate the average latency and the average throughput. Latency is measured as the end-to-end latency as observed by the loadgenerator which receives feedback back from the reporter. Throughput is measured in the reporter component. 

\textit{Results.}
We display our results in Figure~\ref{fig:latency} and \ref{fig:throughput}.
When benchmarking the non-replicated SA as baseline, then we observe throughput up to $233$ op/s (however at a latency of $877$ ms). Generally, we observe that latency increases fast when increasing the workload in the system (see Figure~\ref{fig:latency}).
The system with replicated processors displays a noticeable drop in peak performance: The saturation point is $104$ op/s; after that only latency increases with higher workloads. Because of our observation from the last experiment, we think that this performance loss is not induced by the \bftsmart{} middleware, but rather by the overhead of consolidating outputs as the consolidator runs in \sorrir's Node.js thread (which displays $100\%$ CPU core utilization during the experiment). However, at the level of an acceptable system latency (below $100$ ms) both systems achieve comparable throughput.

\subsection{Leader Failure} \label{eval:leader-failure}
In our third experiment we observe the impact of a replica failure in the replicated system. For this purpose, we crash the leader, which leads to a re-election after some timeout. Our expectation is that 
the system resumes working correctly.

\textit{Measurements.} We use the same benchmark application as in the previous experiment and employ a workload of roughly $50$ op/s. During the experiment, roughly at time $t_{crash}=18$~s we terminate both the Node and Java process at the leader replica, thus stimulating a crash. 

\textit{Results.} 
Figure~\ref{fig:faulty-leader:latency} and \ref{fig:faulty-leader:throughput} show the impact of this crash on system performance.
During the re-election time of slightly above $4$ s, the throughput drops to zero. As expected, we observe system performance bouncing back  after a new leader is elected (roughly at $t_{new} = 22.8$ s). This also leads to an observable peak latency of $4456$~ms (see Fig.~\ref{fig:faulty-leader:latency}).
During this process no events were lost, they were only delayed.

\subsection{Discussion of Results}
From the obtained results
we got the following insights: First, employing \bftsmart{} as replication middleware was an efficient design decision. Due to its implementation in Java, it is easily portable and works on a variety of devices. On the  ARM-Cortex-A72-based Raspberry Pi 4B, we reached competitive throughput levels of over 2.8 kop/s even with 1~kb payload. 
Second, the replication overhead on throughput on the \sorrir{} component level is noticeable (see Figure~\ref{fig:throughput}), which is caused by the \sorrir{} component-based implementation of consolidation, that seems to bottleneck the system. However, consolidating events in the \sorrir{} framework was a design decision that allows (1)~developers to implement customizable consolidation behavior, e.g., relaxing strict consistency of outputs and (2)~flexible instantiation of consolidators, even on devices that do not run a Java environment, or can not establish a TCP connections as they might be wired over Bluetooth. This design lowers the entry barriers for IoT components to interact with replicated components. For low system latency (below $100$ ms) replicated, and non-replicated systems achieve comparable throughput, that we consider practical for many IoT applications. 
Third, we tested the effectiveness of our replication method to tolerate faults, concluding that the replicated system can swiftly return back to its intended behavior.

\section{Conclusion} \label{conclusion}
BFT SMR allows IoT infrastructures to tolerate random data corruption (IoT devices may lack error correcting memory) or mask malicious behavior if components are deployed in dispersed (and potentially untrustworthy) administrative domains.
However, using existing SMR libraries to develop an IoT-fit replication-enriched software architecture is far from trivial.

We show how to automate the whole process of integrating BFT SMR into a component-based IoT execution environment. As solution, we introduce a building-block principle for SMR which flexibly glues replication functionalities to IoT components.

Our solution 
allows for customizable BFT replication mechanisms to be seamlessly integrated in component-based, event-driven execution models like \sorrir. Furthermore, our solution also addresses the automated orchestration and deployment of component replicas on the available hardware landscape.
Experiments conducted on a testbed of Raspberry Pis demonstrate the practicability of our solution.



\bibliographystyle{IEEEtran}
\bibliography{IEEEabrv,bibliography}

\end{document}

%% file: diagrams/ordering.tex
 \begin{tikzpicture}
    \begin{axis}[
width= 5.3cm,
height=4.3cm,
font= \small, 
    xlabel={Throughput [kOp/s]},
    ylabel={Latency [ms]},
    xmin=0, xmax=8,
    ymin=0, ymax=60,
    xtick={0,1,2,3,4,5,6,7,8},
    ytick={0, 10, 20, 30, 40, 50 , 60},
    legend pos=south east,
    legend columns = 2,
    legend style={at={(1 , 1.02)}},
    legend cell align={left},
    ymajorgrids=true,
    xmajorgrids=true,
    grid style=dashed,
]

 \addplot[
     color=JungleGreen,
     mark=diamond,
     ]
     table [x=throughput,y=latency] {data/bftsmart_0.txt};
    
\addplot[
    color=Peach,
    mark=*,
    ]
    table [x=throughput,y=latency] {data/bftsmart_1k.txt};
         \node[right,black!60] at (axis cs:1,5) {{\footnotesize $\longrightarrow$ increasing workload}};
         \node[right,black!60] at (axis cs:1.4,5.1)[circle,fill,inner sep=0.7pt] {};
            \node[right,black!60] at (axis cs:1.75,5.1)[circle,fill,inner sep=0.7pt] {};
   \legend{0 bytes, 1~kb}
\end{axis}
\end{tikzpicture} 
\vskip -0.1 cm

%% file: diagrams/latency.tex
 \begin{tikzpicture}
    \begin{axis}[
width= 5.3cm,
height=4.3cm,
font= \small, 
    xlabel={Backlog Size},
    ylabel={Latency [ms]},
    xmin=0, xmax=200,
    ymin=0, ymax=2100,
    xtick={0,25,50,100,200},
    ytick={0,250, 500, 1000, 2000},
    legend pos=south east,
    legend columns = 2,
    legend style={at={(1 ,1.02)}},
    legend cell align={left},
    ymajorgrids=true,
    xmajorgrids=true,
    grid style=dashed,
]

\addplot[
    color=NavyBlue,
    mark=triangle,
    ]
    table [x=backlog,y=latency] {data/latency.txt};
    
    \addplot[
    color=BrickRed,
    mark=square,
    ]
    table [x=backlog,y=latency] {data/latency_replicated.txt};
   \legend{Base, Replicated}
\end{axis}
\end{tikzpicture} 
\vskip -0.1 cm

%% file: diagrams/throughput.tex
 \begin{tikzpicture}
    \begin{axis}[
width= 5.3cm,
height=4.3cm,
font= \small, 
    xlabel={Backlog Size},
    ylabel={Throughput [Ops/s]},
    xmin=0, xmax=200,
    ymin=0, ymax=250,
    xtick={0,25,50,100,200},
    ytick={0,50,100,150,200,250},
    legend pos=south east,
    legend columns = 2,
    legend style={at={(1, 1.02)}},
    legend cell align={left},
    ymajorgrids=true,
    xmajorgrids=true,
    grid style=dashed,
]

\addplot[
    color=NavyBlue,
    mark=triangle,
    ]
    table [x=backlog,y=throughput] {data/throughput.txt};
\addplot[
    color=BrickRed,
    mark=square,
    ]
    table [x=backlog,y=throughput] {data/throughput_replicated.txt};
   
   \legend{Base, Replicated}
\end{axis}
\end{tikzpicture} 
\vskip -0.1 cm

%% file: diagrams/faulty_leader_throughput.tex
 \begin{tikzpicture}
    \begin{axis}[
width= 4.5cm,
height=4cm,
font= \footnotesize, 
    xlabel={Time (s)},
    ylabel={Throughput [Ops/s]},
    xmin=0, xmax=40,
    ymin=0, ymax=100,
    xtick={0, 10, 20, 30, 40},
    ytick={0,25, 50, 75, 100,150,200,250},
    legend pos=south east,
    legend columns = 2,
    legend style={at={(0.98, 0.8)}},
    legend cell align={left},
    ymajorgrids=true,
    xmajorgrids=true,
    grid style=dashed,
]

\addplot[
    color=RoyalPurple,
    mark=+,
    ]
    table [x=timestamp,y=throughput] {data/faulty_leader_throughput.txt};

   \legend{Throughput}
\end{axis}
\end{tikzpicture} 
\vskip -0.1 cm

%% file: diagrams/faulty_leader_latencies.tex
 \begin{tikzpicture}
    \begin{axis}[
width= 4.5cm,
height=4cm,
font= \footnotesize,
    xlabel={Time (s)},
    ylabel={Latency [ms]},
    xmin=0, xmax=40,
    ymin=0, ymax=200,
    xtick={0, 10, 20, 30, 40},
    ytick={0, 50, 100, 150, 200, 500, 1000},
    legend pos=south east,
    legend columns = 2,
    legend style={at={(0.98, 0.8)}},
    legend cell align={left},
    ymajorgrids=true,
    xmajorgrids=true,
    grid style=dashed,
]

\addplot[
    color=Black,
    mark=.,
    ]
    table [x=timestamp,y=latency] {data/faulty_leader_latencies.txt};

   \legend{Latency}
\end{axis}
\end{tikzpicture} 
\vskip -0.1 cm